%% file: main.tex
\documentclass[%
 reprint,
 superscriptaddress,
 amsmath,amssymb,
 aps,
]{revtex4-2}

\usepackage{graphicx}
\usepackage{dcolumn}
\usepackage{bm}
\usepackage{comment}

\newcommand{\newfigureflexh}[5]{
  \begin{figure}[h]
    \centering
    \includegraphics[width=#5]{#1}
    \caption[#2]{\textbf{#2.} #3}
    \label{#4}
  \end{figure}
}

\newcommand{\newfigureWIDEtop}[5]{
  \begin{figure*}[!t]
    \centering
    \includegraphics[width=#5]{#1}
    \caption[#2]{\textbf{#2.} #3}
    \label{#4}
  \end{figure*}
}

\newcommand{\newfigureWIDEtoph}[5]{
  \begin{figure*}[!ht]
    \centering
    \includegraphics[width=#5]{#1}
    \caption[#2]{\textbf{#2.} #3}
    \label{#4}
  \end{figure*}
}

\newcommand{\newfigureWIDEboth}[5]{
  \onecolumngrid 
  
  \begin{figure}[!hb]
    \centering
    \includegraphics[width=#5]{#1}
    \caption[#2]{\textbf{#2.} #3}
    \label{#4}
  \end{figure}
  \twocolumngrid 
}

\begin{document}

\preprint{APS/123-QED}

\title{Emergent properties of collective gene expression patterns in multicellular systems}   

\author{Matthew Smart}
\email{msmart@physics.utoronto.ca}
\affiliation{Department of Physics, University of Toronto, Canada}

\author{Anton Zilman}%
\email{zilmana@physics.utoronto.ca}
\affiliation{Department of Physics, University of Toronto, Canada}
\affiliation{Institute for Biomedical Engineering, University of Toronto, Canada}

\date{\today}

\begin{abstract}
Multicellular organisms comprise a diverse collection of stable tissues built from different cell types. 
It remains unclear how large numbers of interacting cells can precisely coordinate their gene expression during tissue self-organization. 
We develop a generalized model of multicellular gene expression that includes intracellular and intercellular gene interactions in tissue-like collectives. 
We show that tuning the intercellular signaling strength results in a cascade of transitions from single-cell autonomy towards different self-organized collective states. 
Despite an enormous number of possible tissue states, signaling tends to stabilize a small number of compositionally and spatially simple tissue types even for disordered interaction networks. 
Statistical properties of the stable phenotypes are preserved for different interaction networks and initial conditions. 
These results provide a theoretical framework to study how collections of cells in distinct organisms robustly self-organize into relatively simple tissues even for complex interaction networks mediated by large numbers of different molecules, and how different stable tissues are maintained in homeostasis and disease. 
The close alignment between this theoretical model of tissue self-organization and modern sequencing techniques, particularly spatial transcriptomics, will enable future applications in broad biological contexts. 
\end{abstract}

\maketitle


\section{Introduction}
\label{multicell:intro}
Tissue self-organization, enabled by the formation of distinct patterns of collective gene expression, is a fundamental aspect of multicellular life. 
It remains unclear how very large numbers of interacting cells can precisely coordinate their phenotypes to ensure proper development, homeostasis, and response to environmental challenges.
A key turning point in our understanding of gene regulation within single cells and tissues was the discovery that cells can be experimentally ``reprogrammed" to an undifferentiated state by turning specific genes on or off \cite{Takahashi2006}. 
It is now widely recognized that cellular phenotypes are highly plastic not only \textit{in vitro} \cite{Takahashi2006, Xu2015} but also \textit{in vivo}. 
Cell type transitions are critical for proper tissue homeostasis and wound healing \cite{DeSousaeMelo2019, Guerrero-Juarez2019, Plikus2017, Sinha2018}, leading to a very dynamic picture of tissue self-organization. 
Accordingly, disruption of the self-organized tissue state through gene regulatory perturbations is implicated in the pathogenesis of many diseases, most notably cancer \cite{Giroux2017, Means2005, Smart2021, Yuan2019}. 

A major challenge for understanding tissue self-organization, in light of this plasticity, is the complexity of gene regulation. Vast networks of interacting regulatory molecules underpin the establishment and maintenance of the various tissue states that compose developed organisms. These genetically encoded networks are typically quite different from one organism to the next, emphasizing the need for generalizable models which can provide the synthesis of this diversity towards general principles. Motivated by the regularity of cell fate specification during development, Waddington presciently put forward the concept of an ``epigenetic landscape" in the 1950s \cite{Waddington1957}. This seminal concept has been mathematically expanded as a result of experimental advances \cite{Chang2008, Huang2012, Huang2005, Rand2021, Saez2021, Wang2011}. In particular, modern experimental techniques such as single-cell RNA sequencing (scRNA-seq) have generated enormous amounts of transcriptomic data \cite{Han2018, Karlsson2021, Regev2017, Rozenblatt-Rosen2017}. Clusters in the data represent stable regions in single-cell gene expression space, and can therefore be thought of as dynamical attractors -- the valleys of the epigenetic landscape \cite{Jang2017}. The expression signatures of certain clusters tend to align with known sets of marker genes expressed by \textit{in vivo} cell types \cite{Han2018, Jang2017, Karlsson2021}; this has led to the Human Cell Atlas (HCA) project \cite{Karlsson2021, Regev2017, Rozenblatt-Rosen2017} and related data-driven efforts to characterize the cell types composition of human tissue from development to adulthood and in disease. The large number of observed clusters in the scRNA-seq data indicates that the dynamical system governing single-cell state is nonlinear and high-dimensional. 

It remains an open question how tissues are able to organize into the numerous and diverse architectures that make up multicellular organisms from the building blocks of genetically identical cells. Due to significant limitations in experimental knowledge preceding single-cell techniques, previous approaches to study tissue self-organization focused on either low-dimensional theoretical models, which fall into several categories, or on large systems biology models for specific tissues in specific organisms. 

One important category of the low-dimensional models, which includes classical work \cite{Turing1952} and more recent variants \cite{Murray2002}, explains how spatiotemporal patterns can arise from reaction-diffusion dynamics of morphogens in a cellular ``continuum". This type of approach captures certain spatial features of tissue self-organization, but does not normally describe gene regulatory effects within single cells, such as cellular autonomy and heterogeneity in the absence of morphogens. This limitation has been partially addressed by a different class of approaches that use cellular automata to integrate both single-cell and tissue level effects \cite{Dang2020, Maire2015, Olimpio2018}. However, due to their relatively ad-hoc nature it is not clear how to systematically extend them to larger gene networks associated with scRNA-seq data. To sidestep this fundamental challenge, there is growing interest in a third class of models, so-called ``gene-free" approaches, which implicitly describe complex gene regulation within interacting single cells by working with an abstract phenotype space inspired by the Waddington landscape concept \cite{Camacho-Aguilar2021, Corson2017, Corson2012, Corson2017a, Rand2021, Saez2021}. However, it remains unclear how to relate these abstract variables to the underlying -- and experimentally measurable -- gene activity, and how to incorporate additional cell types or signaling pathways once the model is constructed. 

In contrast to the various types of low-dimensional theoretical frameworks, systems biology approaches have been widely used to model gene regulation within cells in different organisms \cite{Klipp2011}, but parameterization difficulties inherent to many such models make it difficult to draw generalizations from their applications to specific systems. More comprehensive computational approaches which additionally consider cell proliferation and other factors have recently been reviewed \cite{Osborne2017}. Overall, previous works have commonly focused on analytically studying cases with a small number of interacting genes, or use large systems biology models fine-tuned to study specific experimental systems. 

Despite significant advances, it remains unclear how the diverse collections of tissues that make up developed organisms can be generated and maintained using genetically pre-determined rules. The key puzzle is the tendency of multicellular systems to self-organize into stable configurations which are spatially and compositionally simple, containing only a small fraction of the possible phenotypic richness suggested by the combinatorics of collective gene expression. To address this, it is critical to integrate the high-dimensionality of gene expression at the single-cell level with cell-cell interactions which can channel this high-dimensionality into relatively low-complexity tissue states. As a representative example, mammalian cells exhibit over a hundred stable phenotypic states (cell types), and have on the order of $10^4$ protein-coding genes, with ${\sim}10^3$ identified as transcription factors (genes that regulate the expression of other genes) \cite{Babu2004}. Incorporating the high-dimensionality of transcriptional regulation -- which can now be interrogated experimentally using modern sequencing techniques -- provides a path towards addressing questions surrounding the assembly of diverse tissues as well as cell type plasticity across different tissue microenvironments. 

To tackle this key problem, we use models from statistical physics, which are naturally suited to describe how interactions between many microscopic degrees of freedom within and between the cells lead to the formation of stable macroscopic tissue states. To align our work with the conventional definitions of cell type introduced above, we focus on minimal models which can encode a large set of high-dimensional, binarized gene expression patterns as dynamical attractors. These attractors will morph in the presence of cell-cell signaling, reminiscent of cellular plasticity \textit{in vivo}, thereby facilitating the self-organization of diverse tissue configurations. We specifically employ a type of spin glass \cite{Amit1989, Hopfield1982} which has been used to describe single-cell reprogramming experiments \cite{Lang2014, Pusuluri2018}, generalizing it here to describe multistability in cellular collectives. 

In this paper, we present a model of multicellular gene expression that couples the transcriptomic states of interacting cells in a systematic, tunable manner. The model allows us to investigate the interplay between the intra- and intercellular gene-gene interactions in forming stable collective tissue states in a framework that is amenable to experimental input and verification by single-cell transcriptomics data. We demonstrate how different choices of cell-cell interactions can cause the multicellular system to self-organize into a broad range of collective spatial patterns. To examine how cell-cell interactions control the formation of such patterns, we consider an ensemble of tissues composed of non-interacting cells and tune the strength of signaling. This reveals a rich sequence of transitions in the space of tissue gene expression. In the strong signaling regime, we characterize the distribution of stable tissues and show that it may be partitioned into a relatively small number of tissue types, offering insight into the assembly of diverse tissue configurations from the coordination of genetically identical but phenotypically plastic cells. Our results suggest that several statistical properties of this distribution are invariant under different realizations of the random signaling rules, while also shedding light on the puzzling prevalence of compositionally simple tissues in nature despite the complex connectivity of the ``hairball" of signaling pathways that govern them \cite{Levchenko2003}. We conclude with a discussion of these results, which have implications for understanding the self-organization of diverse tissues as well as the plasticity of the cell types which constitute them. Our analysis may also inform our understanding of diseases that disrupt the phenotypic composition of tissues, such as autoimmunity and cancer. 

\section{Model}
\label{multicell:model}
We begin by detailing the spin glass model of multicellular gene expression illustrated in Fig. \ref{multicell:fig1} which describes the expression state of $M$ cells each with $N$ genes. 
First, we introduce the fundamental unit of the model, the single cell. 
We then present the multicell model wherein cells interact on a graph describing their spatial couplings.
Finally, we explain how the cell-cell interactions are incorporated and how the state of each cell is updated. 

\subsection{Single-cell model}
\label{multicell:model_sc}
We assume that the phenotype of a single cell is defined by its gene expression pattern, where each gene is in a binary ``on" or ``off" state. 
This simplifying assumption follows Kauffman’s classical work \cite{Kauffman1969}. 
The state of the cell is then given by an $N$-dimensional vector $\mathbf{s} \in \{+1, -1 \}^N$ where $N$ is the number of genes and $s_i$ denotes the state of gene $i$. 

Stable single-cell phenotypes (cell types) are represented as attractors of the biological gene regulatory network \cite{Mojtahedi2016}. 
The transcriptome of a given cell type $\mu$ is denoted by $\boldsymbol{\xi}^{\mu}$ (as an $N$-dimensional binary vector), and ongoing sequencing efforts have identified the stable transcriptomic signatures of a large set of $p$ cell types $\{\boldsymbol{\xi}^{\mu}\}_{\mu =1}^p$. 
The stability of each observed cell type $\boldsymbol{\xi}^{\mu}$ is a key constraint for a candidate model of gene expression dynamics.

Hopfield networks (HN) \cite{Amit1989, Amit1985, Hopfield1982} provide a tractable minimal model to encode such attractors. 
Mathematically, an HN is a form of Ising spin glass and is defined by the Hamiltonian (defining a pseudo-energy landscape)

\begin{equation}
\label{multicell:eq1}
H_0(\mathbf{s})= -\frac{1}{2} \mathbf{s}^\mathsf{T} \mathbf{J} \mathbf{s} - \mathbf{h}^\mathsf{T} \mathbf{s}
\end{equation}

where $\mathbf{h}$ is an applied field on each gene and the gene-gene interactions $\mathbf{J}$ are chosen so that each pattern $\boldsymbol{\xi}^{\mu}$ is a global minimum of Eq. (\ref{multicell:eq1}) in the absence of an external field. 
Choosing $\mathbf{J}$ based on a rule to encode a set of patterns as minima is commonly called pattern storage in the HN literature. 

The pseudo-energy landscape of the spin model reflects the Waddington landscape of the  gene-gene interaction network, with local minima representing the stable configurations of gene expression corresponding to the biologically observed single-cell phenotypes.

Because the gene expression vectors associated with cell types are typically correlated (i.e. non-orthogonal), we use the projection rule 
\cite{Kanter1987, Personnaz1986} for pattern storage as in ref. \cite{Lang2014}. 
Given an $N \times p$ matrix of cell type transcriptomes 
$\boldsymbol{\xi}=[\boldsymbol{\xi}^{1} \:\:\, \boldsymbol{\xi}^{2} \: \ldots \: \boldsymbol{\xi}^{p}]$, 
the projection rule for storing the $p$ patterns is
\begin{equation}
\label{multicell:eq2}
\mathbf{J}^{\textrm{proj}}= \boldsymbol{\xi} (\boldsymbol{\xi}^\mathsf{T} \boldsymbol{\xi})^{-1} \boldsymbol{\xi}^\mathsf{T},
\end{equation}
which specifies an $N \times N$ matrix of gene-gene interactions. As in \cite{Kanter1987} we set the diagonal elements $J_{ii}$ to zero.

Minima of the Hamiltonian Eq. (\ref{multicell:eq1}) correspond to stable single-cell gene expression states. 
In the absence of noise or external signals $\mathbf{h}$, a given cell state $\mathbf{s} \in \{+1, -1 \}^N$ will decrease its energy $H_0(\mathbf{s})$ until it reaches a local minimum $\mathbf{s}^*$ of $H_0$. 
The projection rule Eq. (\ref{multicell:eq2}) ensures each cell type $\boldsymbol{\xi}^{\mu}$ is a global minimum of $H_0$ \cite{Kanter1987, Personnaz1986}. 

This framework has been shown to recapitulate aspects of \textit{in vitro} cellular reprogramming in single cells \cite{Lang2014, Pusuluri2018}. 
HNs have also been used in a variety of other biological contexts \cite{Fard2016, Fard2017, Guo2017, Kramer2021, Maetschke2014, Szedlak2017}. 
However, these works did not account for the cell-cell interactions within tissue, which play an essential role in regulating tissue structure and composition. 
To describe multicellular systems, we next consider how cells may influence one another’s gene expression. 

\subsection{Multicellular model}
\label{multicell:model_mc}
To treat a multicellular system of $M$ cells, we extend Eq. (\ref{multicell:eq1}) by adding intercellular interaction terms. 
The Hamiltonian for the collection of $M$ cells, as a set of $N$-dimensional gene expression vectors $\{\mathbf{s}^a \}_{a=1}^M$, is 
\begin{equation}
\label{multicell:eq3}
H(\{ \mathbf{s}^a \})= \sum_a H_0(\mathbf{s}^a) + \gamma \sum_a \sum_b A^{ab} f(\mathbf{s}^a, \mathbf{s}^b)
\end{equation}
where the summation is over individual cells. 
The first sum describes the $M$ individual cells, while the second sum contains the cell-cell couplings with a symmetric functional form $f(\mathbf{x},\mathbf{y})=f(\mathbf{y},\mathbf{x})$. 
Biologically, these cell-cell interactions can be mediated by a variety of factors, and we detail a particular form for the coupling inspired by ligand-receptor signaling in the following subsection. 
The spatial adjacency matrix $\mathbf{A}$ defines which cells are interacting, with $A^{ab}=1$ if $a$ and $b$ are neighbors (interact) and $0$ otherwise (Fig. \ref{multicell:fig1}(a)). 

The overall interaction strength is quantified by the global parameter $\gamma \ge 0$. 
As $\gamma \rightarrow 0$ the tissue acts as a collection of $M$ independent, non-interacting cells. 
Beyond a certain threshold $\gamma >0$ the system may exhibit emergent multicellular behavior, such as signaling dependent cell types or collective spatial patterns.

\newfigureWIDEtoph{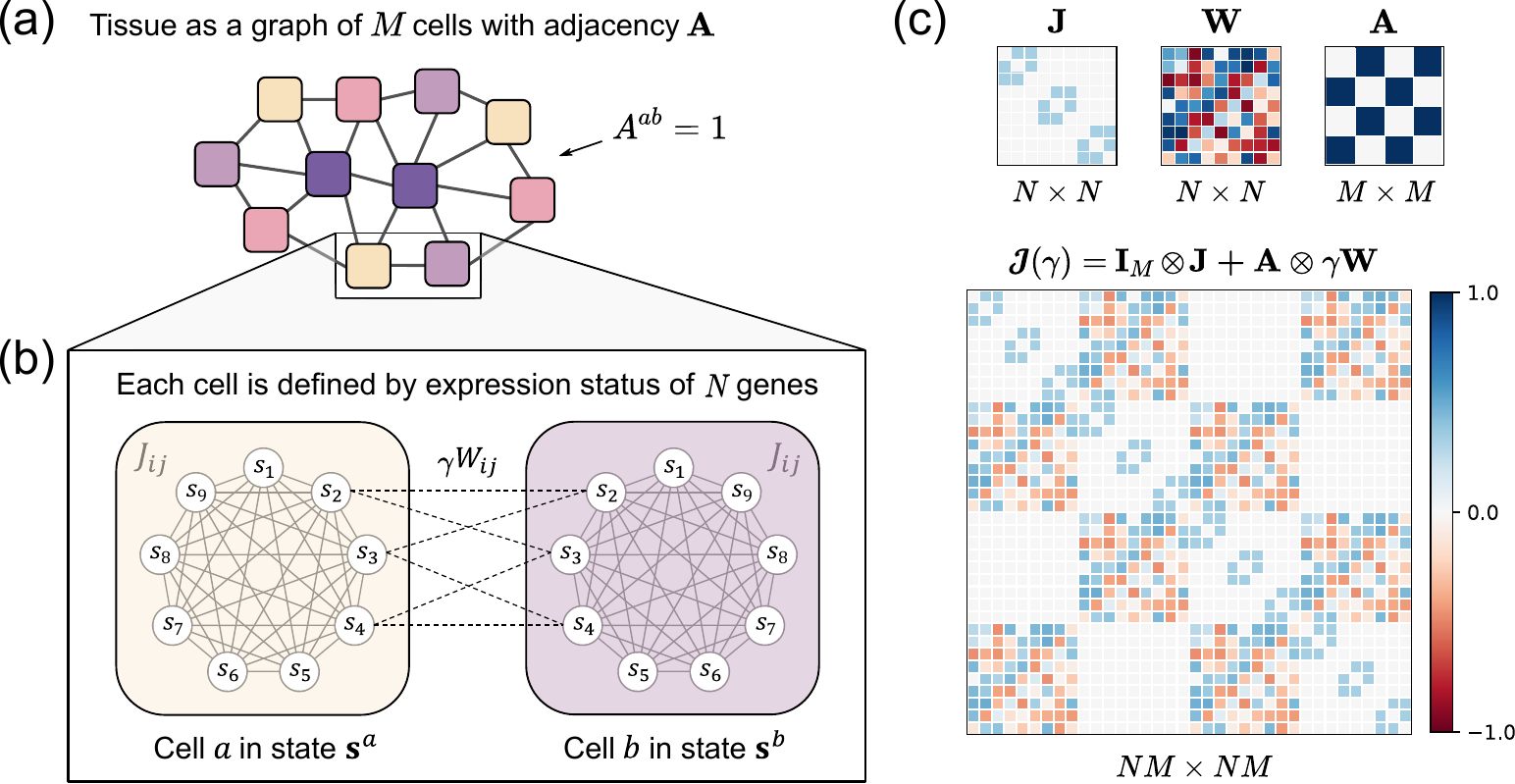}
  {Multicellular model}
  {(a) Biological tissue consisting of $M$ interacting cells is represented by a graph.
  Graph edges are defined by the adjacency matrix $\mathbf{A}$ with $A^{ab}=1$ if cells $a,b$ interact, $0$ otherwise.
  (b) Single-cell state is defined by the expression status of $N$ genes, $s_i \in \{+1,-1\},i=1,\ldots,N$. 
  We concisely denote the state of cell $a$ by the $N$-dimensional vector $\mathbf{s}^a$. 
  Every cell has the same intracellular gene regulatory rules specified by an $N \times N$ matrix $J_{ij}$. 
  Adjacent cells interact according to a second $N \times N$ matrix $W_{ij}$ which describes the effect that gene $j$ in a ``sender" cell has on gene $i$ in a ``recipient" cell. 
  The cell-cell interaction strength $\gamma$ tunes the magnitude of $\mathbf{W}$ relative to $\mathbf{J}$. 
  (c) The interaction matrix for tissue gene expression, denoted by $\bm{\mathcal{J}}(\gamma)$ in Eq. (\ref{multicell:eq2}), defines interactions between all $NM$ genes in the multicellular system. To illustrate its construction, we show $\bm{\mathcal{J}}(\gamma=0.5)$ for a particular choice of $\mathbf{J}$,$\mathbf{W}$, and $\mathbf{A}$.}
  {multicell:fig1}{0.8\linewidth}

For simplicity, we use an adjacency matrix corresponding to next-nearest-neighbor interactions on a square lattice. 
This choice is biologically inspired by cells interacting over relatively short distances (e.g. Notch signaling \cite{Corson2017a}). 
While we focus on the two-dimensional case for ease of visualization, an arbitrary cell-cell interaction matrix can be chosen in principle. 
Likewise, the interaction strength $\gamma$ could also be distance-dependent instead of constant. 

\subsection{Cell-cell interactions}
\label{multicell:model_intxn}
Many genes participate in cell-cell interactions by generating, sensing, or transducing signals that are sent between cells. 
For example, a cell may express and secrete signaling molecules (ligands) into the surrounding environment, which in turn affects the gene expression of neighboring cells that sense and respond to these signals. 
Each ligand may influence the expression of many genes in a target cell resulting in an intercellular gene-gene interaction network. 
We represent this network of sender-recipient signaling interactions via an $N \times N$ matrix $\mathbf{W}$, where $W_{ij}$ represents the effect gene $j$ (in a sender cell) has on gene $i$ (in a recipient cell). 
A complete empirical $\mathbf{W}$ is not yet available from the biological data and so we will focus on the case of randomly sampled $\mathbf{W}$. 

The cell-cell signaling matrix $\mathbf{W}$ effectively couples the gene regulatory networks of neighboring cells, depicted graphically in Fig. \ref{multicell:fig1}(b). 
We denote the ``signaling field" which cell $b$ exerts on a neighboring cell $a$ by 
$\mathbf{h}^{ab}=\gamma \mathbf{W} \mathbf{s}^b$, 
which acts as an applied field on the single-cell Hamiltonian for cell $a$, $H_0(\mathbf{s}^a)$ (Eq. (\ref{multicell:eq1})). 
Summing over all neighbors gives the total applied field that the tissue collectively exerts on cell $a$, 
\begin{equation}
\label{multicell:eq4}
\mathbf{h}^{a}=\gamma \mathbf{W} \sum_b A^{ab} \mathbf{s}^b.
\end{equation}

Note that this collective applied field is not static; it changes with the state of the tissue. This aspect is essential for the self-organizing properties of the model. 

The collective applied field Eq. (\ref{multicell:eq4}) we have considered for our initial investigation depends only on the state of the sender cells. In principle, this function could also depend on the state of the recipient cell. 
For example, a cell that is not expressing certain receptors will not be able to sense and respond to the corresponding ligands. 
This more complex signaling form will be explored in future work. 

When both $\mathbf{J}$ and $\mathbf{W}$ are symmetric, the collective behavior can be studied from a statistical mechanics perspective through the multicellular Hamiltonian introduced above. 
The choice of collective applied field in Eq. (\ref{multicell:eq4}) corresponds to choosing 
$f(\mathbf{s}^a,\mathbf{s}^b) = - \frac{1}{2}{\mathbf{s}^a}^\mathsf{T} \mathbf{W} \mathbf{s}^b$ 
for the interaction terms in Eq. (\ref{multicell:eq3}). 
The parameterized multicellular Hamiltonian is 
\begin{equation}
\label{multicell:eq5}
H(\{ \mathbf{s}^a \}) =
- \frac{1}{2} \sum_a {\mathbf{s}^a}^\mathsf{T} \mathbf{J} \mathbf{s}^a
- \frac{\gamma}{2} \sum_a \sum_b A^{ab} {\mathbf{s}^a}^\mathsf{T} \mathbf{W} \mathbf{s}^b .
\end{equation}

Equation (\ref{multicell:eq5}) is compactly expressed as 
$H(\mathbf{x},\gamma) = -\frac{1}{2} \mathbf{x}^\mathsf{T} \bm{\mathcal{J}} (\gamma) \mathbf{x}$, 
where $x$ is an $NM$ vector of each cell’s transcriptome concatenated, and the $NM \times NM$ interaction matrix $\bm{\mathcal{J}}(\gamma)$ has a block form, \begin{equation}
\label{multicell:eq6}
\bm{\mathcal{J}}(\gamma)= \mathbf{I}_M \otimes \mathbf{J} + \mathbf{A} \otimes \gamma \mathbf{W},
\end{equation}
with $\otimes$ denoting the Kronecker product for constructing block matrices and $\mathbf{I}_M$ the $M \times M$ identity matrix. 
Each term represents a distinct layer of gene regulation: the diagonal blocks correspond to the intracellular gene interactions $\mathbf{J}$, whereas the off-diagonal blocks correspond to the intercellular signaling matrix $\gamma \mathbf{W}$ tiled according to the adjacency matrix $\mathbf{A}$ (see Fig. \ref{multicell:fig1}(c)). 
The cell-cell interaction strength $\gamma$ tunes the magnitude of $\mathbf{W}$ relative to $\mathbf{J}$ (Fig. \ref{multicell:fig1}(b)).

Various types of block Ising models have been studied in different contexts \cite{Barra2011, Fedele2011, Kirsch2020, Knopfel2020}, but they have largely been restricted to either two cells or to uniform off-diagonal interactions (i.e. constant $W_{ij}$ or simple adjacency $A^{ab}=1-\delta^{ab}$). 
We are focused here on much more general off-diagonal interactions (randomly sampled $W_{ij}$ and structured adjacency matrices; see Results). 
Of note, refs. \cite{Agliari2018, Barra2015, Panchenko2015} studied disordered ``multi-species" block systems with an arbitrary number of cells, but they consider alternative couplings between cells and do not focus on the deterministic limit.

Although in this work we consider discrete gene expression states, we note that related continuous state models known as coupled map lattices \cite{Kaneko1992} have been used to describe lattices of interacting cells \cite{Bignone1993, Garcia-Morales2017, Klevecz1998}.
These works focused on a few genes or underlying cell types, whereas the approach we outline is inherently scalable to many genes and cell types. 

\subsection{Gene expression dynamics}
\label{multicell:model_dyn}
The minima of Eq. (\ref{multicell:eq5}) correspond to stable configurations of the tissue (collective gene expression patterns). To identify them, we use a discrete analog of gradient descent on Eq. (\ref{multicell:eq5}). Our results are also relevant in the case of mild gene expression noise. We present the full stochastic update rule here for completeness, then introduce the deterministic limit. 

We use Glauber dynamics \cite{Amit1989, Glauber1963} as an asynchronous update rule for the single-cell spin glass Eq. (\ref{multicell:eq1}) and its multicell extension Eq. (\ref{multicell:eq5}). 
We emphasize that we are not focused on the dynamics itself, but rather in using it as a tool to sample the steady states which arise for interacting cells. 
For a given cell, a gene is selected at random and updated according to 
\begin{equation}
\label{multicell:eq7}
p(s_i(t+1) \rightarrow 1) = \frac{1}
{1 + \exp{(-2 \beta h_i^{\textrm{total}} )} }
\end{equation}
where $h_i^{\textrm{total}}=\sum_j J_{ij} s_j(t) + h_i$, and $\beta^{-1}$ represents the strength of the gene regulatory noise arising from various sources (analogous to thermal noise). 
The timestep $t$ is expressed in units of single gene updates. 
Note that the mean spin update is $\langle s_i (t+1) \rangle = \tanh{(\beta h_i^{\textrm{total}} )}$.

In the deterministic limit ($\beta \rightarrow \infty$) of Eq. (\ref{multicell:eq7}), the update rule for a single cell becomes 
$\mathbf{s}(t+N)=\textrm{sgn}(\mathbf{J}\mathbf{s}(t)+\mathbf{h})$, where 
$\textrm{sgn}(\cdot)$ is applied element-wise and the $N$ genes are updated in a fixed sequential order.
We fix the sequence of updates in order to ensure a well-defined mapping from an initial condition to a resulting fixed point.
A state is a fixed point of the update rule when $\mathbf{s}=\textrm{sgn}(\mathbf{J s + h})$. 
The sequential update order does not impact whether a state is a fixed point. 
For $\mathbf{h}=\mathbf{0}$, it can be verified that the encoded cell types $\boldsymbol{\xi}^{\mu}$ are fixed points when $\mathbf{J}$ is defined via the projection rule Eq. (\ref{multicell:eq2}). 

For the multicellular model, the deterministic update rule for each cell is 
\begin{equation}
\label{multicell:eq8}
\mathbf{s}^a (t+N) = \textrm{sgn}( 
  \mathbf{J} \mathbf{s}^a (t) + 
  \gamma \mathbf{W} \sum_b A^{ab} \mathbf{s}^b (t) 
)
\end{equation}
where $a \in \{1,\ldots,M \}$. 
Tissue level updates can be expressed compactly using Eq. (\ref{multicell:eq6}) as 
$\mathbf{x}(t+NM) = \textrm{sgn}( \bm{\mathcal{J}} \mathbf{x} (t) )$. 

Very similar systems of equations have been used as continuous-time dynamical systems, most notably in Hopfield’s classical work on associative memory \cite{Hopfield1984}. 
They take the form $\frac{\mathrm{d} u_i}{\mathrm{d}t} = -\frac{u_i}{\tau_i} + \sum_j \tanh{(\beta v_j)}$ with $v_i \equiv \sum_j J_{ij} u_j +h_i$, $\tau_i > 0$. 
An analogous system has been applied to scRNA-seq data, where it generated experimentally validated predictions in the context of differentiation \cite{Jang2017}. 
Interestingly, such systems have also been used as recurrent neural networks which may be trained to reproduce time series from other dynamical systems \cite{Funahashi1993}.

When either of $\mathbf{J}$, $\mathbf{W}$ are asymmetric, the stochastic dynamics no longer satisfy detailed balance. 
In this case, the model is known as an asymmetric kinetic Ising system \cite{Aguilera2021, Mezard2011, Roudi2009}. 
In addition to fixed point attractors, such systems can exhibit oscillatory behavior which is necessary to describe phenomena such as the cell cycle or spatiotemporal patterns. 
This out-of-equilibrium dynamics will be investigated in future work. 

\subsection{Low-dimensional system with three cell types and nine genes}
\label{multicell:model_toy}
For simplicity and to facilitate visualization, we consider a low-dimensional system with $N=9$ genes and $p=3$ encoded single-cell types (shown in Fig. \ref{multicell:fig2}(a)). 
Three cell types is the minimal non-trivial encoding, and nine genes gives a large but tractable space of $512$ transcriptomic states for each cell. The set of cell type gene expression vectors $\{ \boldsymbol{\xi}^{\mu}\}$ determines the intracellular gene regulatory interactions $\mathbf{J}$ through the projection rule Eq. (\ref{multicell:eq2}). 

The single-cell energy landscape Eq. (\ref{multicell:eq1}) is depicted in Fig. \ref{multicell:fig2}(b) for the single-cell types from Fig. \ref{multicell:fig2}(a). 
In addition to the three encoded minima, there are five ``spurious" minima consisting of the negation of each cell type, $-\boldsymbol{\xi}^{\mu}$ (due to spin-flip symmetry of $H(\mathbf{s})$), and the sum of the three types, 
$\pm \mathbf{S} = \pm \frac{1}{3}(\boldsymbol{\xi}^1 + \boldsymbol{\xi}^2 + \boldsymbol{\xi}^3)$. 
Although some spurious minima may be biologically significant \cite{Lang2014}, others are likely not (e.g. the negations $-\boldsymbol{\xi}^{\mu}$).
We assume throughout this paper that $\mathbf{h}=\mathbf{0}$, but we note that certain minima can be eliminated using a global external field.
Briefly, the embedding in Fig. \ref{multicell:fig2}(b) is generated by
computing the Hamming distance between four minima 
$\{\boldsymbol{\xi}^1$, $\boldsymbol{\xi}^2$, $\boldsymbol{\xi}^3$, $\mathbf{S} \}$
and each of the $2^N$ possible gene expression states. 
This results in a matrix $\mathbf{X} \in \mathbb{R}^{4 \times 2^N}$. 
Principle component analysis (PCA) is then applied to $\mathbf{X}$ to obtain a two-dimensional embedding 
$\tilde{ \mathbf{X} } \in \mathbb{R}^{2 \times 2^N}$.
The columns of $\tilde{ \mathbf{X} }$ correspond to embeddings of each gene expression state $\mathbf{s}$, which we plot in three dimensions using $H_0(\mathbf{s})$ as the vertical axis.

In the absence of signaling ($\gamma =0$), the stable multicellular states are composed of prescribed single-cell types $\{\boldsymbol{\xi}^{\mu}\}$. 
To characterize the stable tissues which may self-organize in a given multicellular system for $\gamma >0$, we sample local minima of Eq. (\ref{multicell:eq5}) from arbitrary initial conditions for a given choice of intracellular interactions $\mathbf{J}$, cell-cell interactions $\mathbf{W}$, and adjacency matrix $\mathbf{A}$. 
    
As mentioned previously, the cell-cell interaction matrix $\mathbf{W}$ is not yet fully characterized experimentally. 
In addition to the many genes that have direct signaling roles (e.g. ligands and receptors), even more genes participate indirectly in signaling networks (e.g. as co-factors, downstream sensing molecules, activatable transcription factors, etc.) or as cargo for extracellular vesicles such as exosomes which can shuttle RNA between cells \cite{LoCicero2015}. 
In lieu of definitive data constraining $\mathbf{W}$ (in contrast to the single-cell transcriptomics data constraining $\mathbf{J}$), we consider dense symmetric matrices, sampling the upper triangular elements as $W_{ij} \sim U[-1,1]$ (see e.g. Fig. \ref{multicell:fig1}(c)). 
We consider structured alternatives to this simple choice in the discussion.

\section{Results}
\label{multicell:results}

\subsection{Different intercellular signaling rules stabilize qualitatively distinct tissue types}
\label{multicell:results_A}
We are first interested in assessing the range of possible tissue states which self-organize under different signaling rules. 
In particular, we fix all aspects of the model except for the signaling matrix $\mathbf{W}$ (i.e. $\mathbf{J}$, $\mathbf{A}$, and $\gamma$). 
As described above, $\mathbf{J}$ is set by the choice of encoded single-cell types $\{ \boldsymbol{\xi}^{\mu} \}$ which are stable in the absence of signaling (Figs. \ref{multicell:fig2}(a) and \ref{multicell:fig2}(b)), and $\mathbf{A}$ represents a next-nearest-neighbor square lattice. 
We then sample different realizations of $W_{ij} \sim U[-1,1]$ and identify the tissues which self-organize starting from a fixed initial condition of the tissue gene expression. 
See the preceding Section \ref{multicell:model_toy} for details of the simulated system. 

\onecolumngrid

\newfigureWIDEboth{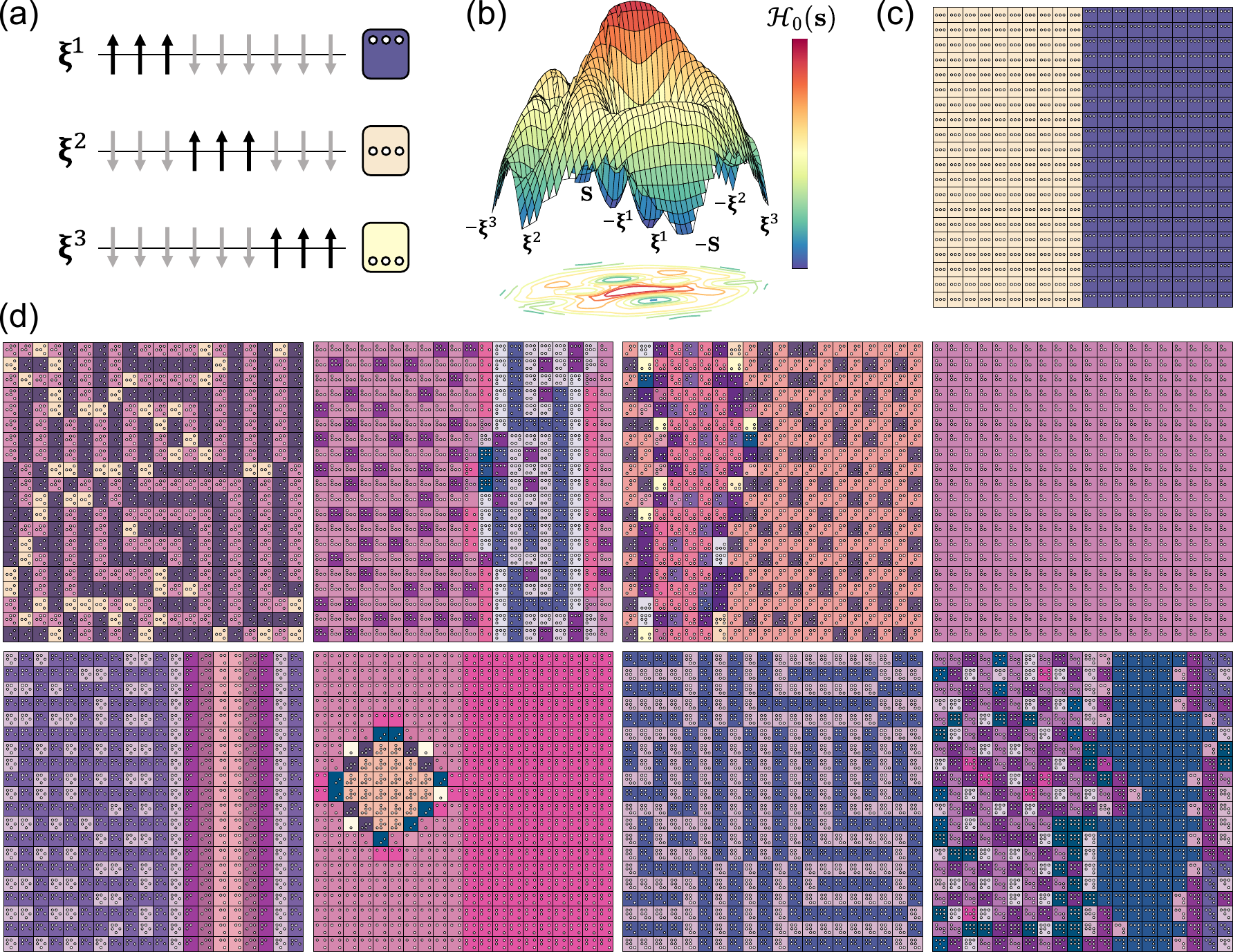}
  {Stable tissue patterns for varying signaling rules}
  {(a) An example system with $N=9$ genes and $p=3$ single-cell types $\{\boldsymbol{\xi}^{\mu}\}_{\mu =1}^3$ defined by their gene expression patterns (arrow up/down denotes on/off). 
  Each cell state is represented by a grid of up to $N$ white dots where the presence (absence) of a dot indicates the associated gene is on (off).
  Additionally, each unique cell state $\mathbf{s} \in 2^N$ is assigned a unique color. 
  The choice of patterns fixes the intracellular rules $\mathbf{J}$ through the projection rule Eq. (\ref{multicell:eq2}). 
  (b) Two-dimensional embedding of the single-cell energy landscape $H_0(\mathbf{s})$ Eq. (\ref{multicell:eq1}). 
  (c) A $400$ cell ($20 \times 20$ grid) lattice initial condition is prepared with $\boldsymbol{\xi}^2$ cells on the left half and $\boldsymbol{\xi}^1$ cells on the right half. 
  (d) Stable tissue patterns reached from the initial condition in panel (c) are found for eight random symmetric signaling rules $W_{ij} \sim U[-1,1]$ with $\gamma=1$.}
  {multicell:fig2}{0.95\linewidth}

\clearpage
\twocolumngrid 

In Fig. \ref{multicell:fig2}(c) we prepare an initial condition of $M=400$ cells arranged on a $20 \times 20$ lattice. 
This choice mimics a 2D sheet consisting of two different cell types: the left half is composed of cells in state $\boldsymbol{\xi}^{2}$ and the right half is composed of cells in state $\boldsymbol{\xi}^{1}$. 
This tissue state evolves according to the regulatory rules, eventually reaching a local minimum of Eq. (\ref{multicell:eq5}). 
Each square in the grid describes a cell at a particular location. 
The gene expression state of each cell is visualized both quantitatively (through the presence/absence of $N=9$ dots corresponding to ``on" genes inside the square as in Fig. \ref{multicell:fig2}(a)) and qualitatively (each state $\mathbf{s} \in 2^N$ is assigned a unique color).

We identify fixed points reached from the initial condition in Fig. \ref{multicell:fig2}(c) for different random interaction matrices W when the signaling strength $\gamma$ is set to $1$. 
Representative examples are displayed in Fig. \ref{multicell:fig2}(d). 
Sufficiently strong interactions can destabilize the encoded single-cell types $\{ \boldsymbol{\xi}^{\mu} \}$. 
The stable configurations reached by different choices of W are quite diverse and can be divided into several ``tissue types". 
These include homogeneous (all cells are in the same state), ordered layers, and labyrinthine (a few states which are spatially interwoven). 
More heterogeneous patterns are possible, as are multi-phasic patterns where different spatial regions of the tissue exhibit different patterns (such as homogeneous in one region and labyrinthine in another). 
The spatial features of the observed patterns also scale with the signaling range (Fig. \ref{multicell:figS1}).

\newfigureflexh{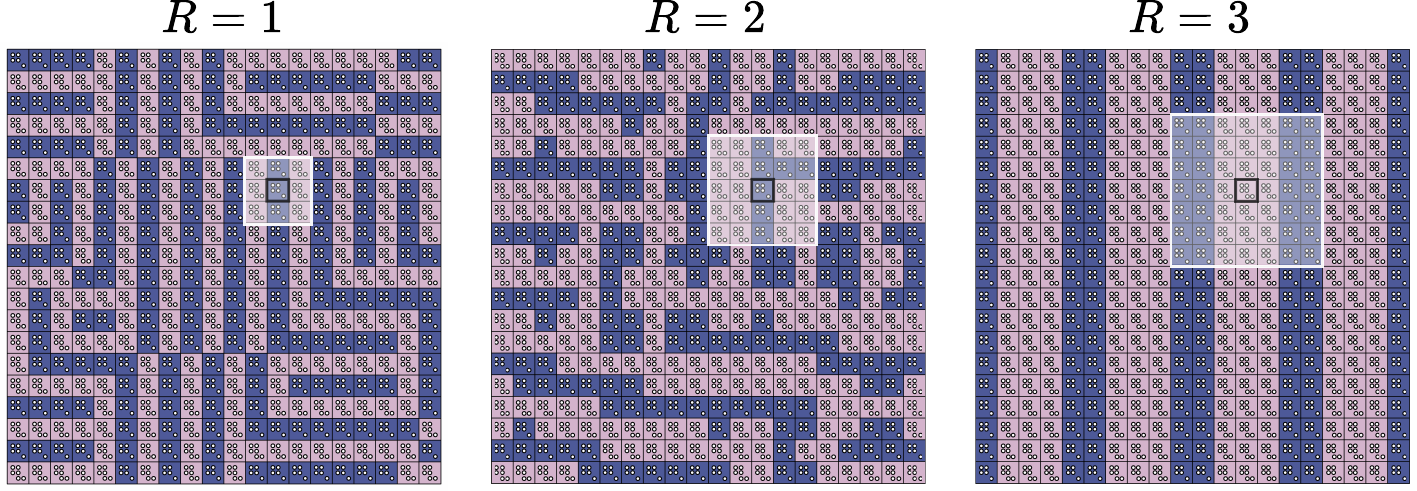}
  {Tissue patterns scale with signal radius}
  {We fix the cell-cell signaling rules $\mathbf{W}$ which stabilized the upper-left tissue state in Fig. \ref{multicell:fig2}(d). 
  Details of the visualization approach are provided in Fig. \ref{multicell:fig2}. 
  Each tissue is generated from the same dual initial condition as in Fig. \ref{multicell:fig2}(c) for $\gamma =1$. 
  The interaction radius $R$, which defines a corresponding adjacency matrix $\mathbf{A}_R$, is indicated above each tissue. 
  The white shaded region denotes the neighborhood which exchanges signals with the indicated cell.}
  {multicell:figS1}{0.9\linewidth}

Overall, different signaling rules $\mathbf{W}$ can cause an arbitrary configuration of single-cells to self-organize into a variety of stable tissue configurations. 
These patterns are maintained through coherent signaling between the cells within their local neighborhoods (Eq. (\ref{multicell:eq4})). 
This tissue self-organization process is important for biological functions such as homeostasis in different organisms but is difficult to characterize experimentally due to the high-dimensionality of gene expression. 
This theoretical approach describes the formation and stabilization of collective gene expression patterns for particular choices of the gene regulatory parameters ($\mathbf{J}$,$\mathbf{W}$,$\gamma$) and spatial organization of cells ($\mathbf{A}$), which may aid understanding of how particular stable states arise both \textit{in vivo} and \textit{in vitro}. 

\subsection{Tuning the signaling strength destabilizes single-cell types, leading to a rich sequence of tissue transitions}
\label{multicell:results_B}
Complex organisms exhibit a broad range of stable tissue configurations. It remains unclear how such diversity arises in the context of genetically predefined intracellular and intercellular regulatory rules (i.e. $\mathbf{J}$, $\mathbf{W}$ are fixed by the genome). In a given organism there are numerous tissue types that are distinguished by their cellular composition and spatial patterning. To carry out different functional roles, such tissue types are often strikingly distinct (e.g. adipose tissue and hepatic lobules). In addition to the macroscopic variation across tissue types, there can also be fine-scale variation in a given type. In this case, the tissues are constructed from the same set of cell states, but there is minor variation in cell number or spatial arrangement (e.g. the fingerprints of identical twins). 

Here we characterize the diversity of stable tissue configurations in the multicellular model when the regulatory parameters $\mathbf{J}$, $\mathbf{W}$, and $\mathbf{A}$ are fixed. We analyze how this diversity emerges as the cell-cell signaling strength $\gamma$ is tuned, as might occur during development or as part of homeostasis and adaptation to environmental pressures. 

As an illustrative example, in Fig. \ref{multicell:fig3}(a) we reconsider the initial condition from Fig. \ref{multicell:fig2}(c) and gradually increase $\gamma$ for a particular choice of cell-cell signaling rules $\mathbf{W}$. We keep the same $\mathbf{J}$, $\mathbf{A}$ as in Fig. \ref{multicell:fig2} but lower the size of the system to a $10 \times 10$ lattice ($M=100$ cells) to facilitate computation and visualization. The initial configuration (Fig. \ref{multicell:fig3}(a), left) is composed of single-cell types that are stable in isolation (i.e. in the absence of signaling), and it therefore remains stable for very mild levels of signaling. However, once $\gamma$ passes a certain threshold, the tissue is destabilized and self-organizes into a different stable configuration. These multicellular patterns are characterized by signaling-dependent single-cell states (i.e. they are maintained by cell-cell interactions). As $\gamma$ increases the tissue undergoes a cascade of such transitions which are punctuated by intervals of stability. 
At $\gamma \gtrsim 1$ the tissue reaches a limiting configuration that no longer changes, which we term the strong signaling regime. 

Fig. \ref{multicell:fig3}(a) displays only a few of the many tissue transitions which are observed from this particular initial condition. While the transition sequence appears quite complex overall, the earliest transitions at low $\gamma$ may be anticipated. 
By definition, a tissue state is stable if all cells $\{ \mathbf{s}^a \}$ present in the tissue are stable, as defined by Eq. (\ref{multicell:eq8}). 
Because the tissue states at $\gamma=0$ consist of only stable single-cell types $\boldsymbol{\xi}^{\mu}$, we can readily enumerate the possible neighborhoods (which are defined by the adjacency matrix $\mathbf{A}$). 
These neighborhoods determine the collective signaling field experienced by a given cell, Eq. (\ref{multicell:eq4}). 
The simplest neighborhood present in the initial condition in Fig. \ref{multicell:fig3}(a) consists of a cell in state $\boldsymbol{\xi}^{\mu}$ surrounded by $z \equiv 8$ cells also in state $\boldsymbol{\xi}^{\mu}$ ($z$ is the coordination number for $\mathbf{A}$). 
We therefore ask: at what $\gamma$ does the fixed point condition 
$\boldsymbol{\xi}^{\mu}=\textrm{sgn}(
\boldsymbol{\xi}^{\mu} + \gamma z \mathbf{W} \boldsymbol{\xi}^{\mu}
)$ 
no longer hold?

In this example, the right half of the lattice is composed of $\boldsymbol{\xi}^{1}$ while the left half is composed of $\boldsymbol{\xi}^{2}$. For $\mu=1$ this criterion gives $\gamma_1^* \approx 0.0231$, whereas for $\mu=2$ it gives $\gamma_2^* \approx 0.0559$. This analysis is reflected in Fig. \ref{multicell:fig3}(a), which shows that by $\gamma =0.024$ the right half is destabilized but not the left, and by $\gamma =0.056$ the left half becomes destabilized as well. A general consequence of this analysis is that the encoded single-cell types $\{ \boldsymbol{\xi}^{\mu} \}_{\mu =1}^p$ are destabilized in a sequential fashion, from which we identify at least $\sim p$ low-$\gamma$ transitions. This heuristic is less useful for describing the many transitions at intermediate levels of $\gamma$ because it requires enumerating all possible neighborhoods and the combinatorics become prohibitive. 

\subsection{Nonlinear dimension reduction reveals the emergence of self-organized tissue types}
\label{multicell:results_C}
To generalize beyond a single initial condition, we numerically investigate the distribution of stable gene expression states (of the entire tissue) reached by an ensemble of initial conditions. 
Specifically, we choose $k=10^4$ random initial conditions $\{\mathbf{x}_i^0 \}_{i=1}^k$ and for each we compute the stable tissue $\mathbf{x}_i$ which self-organizes at a particular value of $\gamma$. 
This mapping generates a large matrix $\mathbf{X}_{\gamma} \in \{+1,-1\}^{NM \times k}$. 
Each column of $\mathbf{X}_{\gamma}$, denoted $\mathbf{x}_i$, is a local minimum of Eq. (\ref{multicell:eq5}). 

Classically, low-dimensional ``order parameters" (e.g., the magnetization $m=N^{-1} \sum_i s_i$ in the case of a ferromagnet) have served as the quantification of the degree of disorder and a way to categorize different macroscopic phases. 
However, the rich space of distinct minima in our systems makes direct identification of order parameters impractical. 
Instead we use an unsupervised nonlinear dimension reduction technique, Uniform Manifold Approximation and Projection (UMAP) \cite{McInnes2018}, to embed and identify the tissue states that belong to the same class based on their phenotypic composition and spatial gene expression patterns. 
Unsupervised learning techniques are increasingly being leveraged to study complex physical systems \cite{Lopez2019, Mehta2019, Wang2018}, and a similar approach to the one taken here has recently been used to visualize high-dimensional energy landscapes in materials science \cite{Shires2021}.

Despite the high-dimensionality of the data, UMAP is able to generate informative two-dimensional embeddings. In Fig. \ref{multicell:fig3} (b) and (c) we display the nonlinear embedding of $\mathbf{X}_{\gamma}$ for several representative values of $\gamma$. Each point in a given panel corresponds to a stable tissue $\mathbf{x}_i$. In Fig. \ref{multicell:fig3}(b) we color these points by $n(\mathbf{x}_i)$, which denotes the number of unique single-cell states that are present in the tissue, whereas Fig. \ref{multicell:fig3}(c) shows the same points colored according to their relative energy, $H(\mathbf{x}_i)$. 

\newfigureflexh{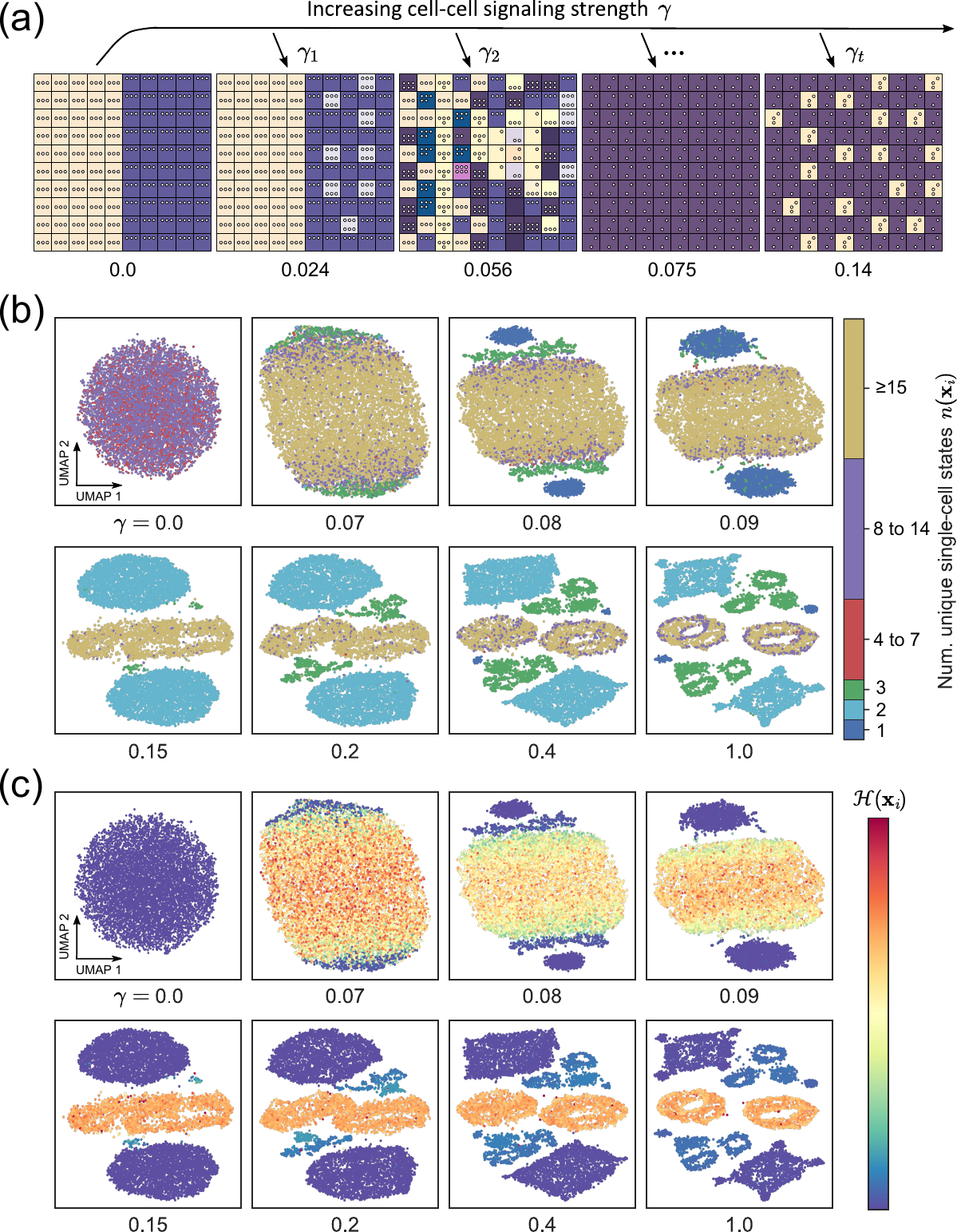}
  {Emergence of clusters of stable tissue patterns through cell-cell signaling}
  {We fix $\mathbf{J}$ as in Fig. \ref{multicell:fig2} and sample a particular set of cell-cell signaling rules $W_{ij} \sim U[-1,1]$. 
  We use the same type of adjacency matrix $\mathbf{A}$ as in Fig. \ref{multicell:fig2} but with a $10 \times 10$ lattice ($M=100$ cells). 
  (a) From a specific tissue initial condition, we increase $\gamma$ and identify the resulting stable configuration (the same initial condition is used for each $\gamma$). 
  Numerous tissue transitions are observed as $\gamma$ is tuned (see Fig. \ref{multicell:fig2} for visualization details). 
  (b) We fix $k=10^4$ random initial conditions and compute the corresponding stable tissue states $\mathbf{x}_i \in \{+1,-1 \}^{NM}$ ($N$ genes, $M$ cells) for eight different values of $\gamma$. 
  For each $\gamma$, this generates a matrix $\mathbf{X} \in \{+1,-1 \}^{NM \times k}$. 
  We embed $\mathbf{X}$ in two dimensions using nonlinear dimension reduction (aligned UMAP \cite{McInnes2018}). 
  Each point $\mathbf{x}_i$ is colored according to the number of unique single-cell states present in the tissue, $n(\mathbf{x}_i)$. (c) Each point in (b) is colored instead by its multicellular energy $H(\mathbf{x}_i)$.}
  {multicell:fig3}{1.0\linewidth}

At $\gamma = 0$, the points are embedded in a relatively unstructured manner. This is expected and suggests that the way we are sampling local minima is not overtly biased. 
A heuristic argument for this observation is as follows: in the absence of interactions ($\gamma = 0$), a tissue state is stable whenever each cell is stable. 
The number of stable single-cell states $n$ includes the $p$ encoded single-cell types $\{ \boldsymbol{\xi}^{\mu} \}$ but can also include additional spurious stable states. For the system with $p=3$ encoded cell types considered here, there are n=8 stable states (Fig. \ref{multicell:fig2}(b)). This gives $n^M$ distinct stable tissue states. When this very large space of $n^M$ minima is sub-sampled at random, one expects little structure in the nearest-neighbor graph. This disorder is ultimately reflected in the embedding.


As $\gamma$ increases this unstructured fine-scale diversity of non-interacting cells in arbitrary spatial arrangements gradually coalesces into a relatively small number of clusters. Interestingly, these clusters have a clear signature in terms of $n(\mathbf{x})$ -- the unique single-cell states that are present in the tissue -- as well as the energy $H(\mathbf{x})$ Eq. (\ref{multicell:eq5}) (as shown in Fig. \ref{multicell:fig3} (b) and (c)). This indicates that the embedding is clustering distinct ``types" of tissue (in analogy to cell types) which maintain some intra-cluster variation. For instance, for $\gamma \geq 0.15$ there are two large clusters which contain tissues composed of just two cell states. 

Overall, tuning the interaction strength $\gamma$ promotes the self-organization of a wide array of stable tissue configurations. At $\gamma = 0$, the model is simply a collection of $M$ non-interacting single-cell types. As $\gamma$ increases, multicellular structures emerge through collective interactions between the cells. We have shown that the low-$\gamma$ transitions are associated with sequential destabilization of the encoded single-cell types, whereas the transitions for higher $\gamma$ are more nuanced. In the next subsection, we show that the unsupervised approach we take here is especially useful for identifying and distinguishing tissue types in the strong signaling regime. 

\subsection{Strong signaling causes the tissue to self-organize into a relatively small number of types}
\label{multicell:results_D}
In the preceding subsection, we applied nonlinear dimension reduction to the stable gene expression patterns $\mathbf{X}_{\gamma}$ reached from an ensemble of random initial conditions. We showed that the self-organized tissue states form several clusters in the low-dimensional space as $\gamma$ is increased. Here we focus on $\mathbf{X}_{\gamma}$ in the limiting regime of strong signaling ($\gamma = 1$). Fig. \ref{multicell:fig4}(a) shows the UMAP embedding of $\mathbf{X}_{\gamma =1}$ colored by the number of unique single-cell states present in the tissue, $n(\mathbf{x}_i)$. Several representative points from each cluster are annotated and visualized in Fig. \ref{multicell:fig4}(b). 

\newfigureWIDEtop{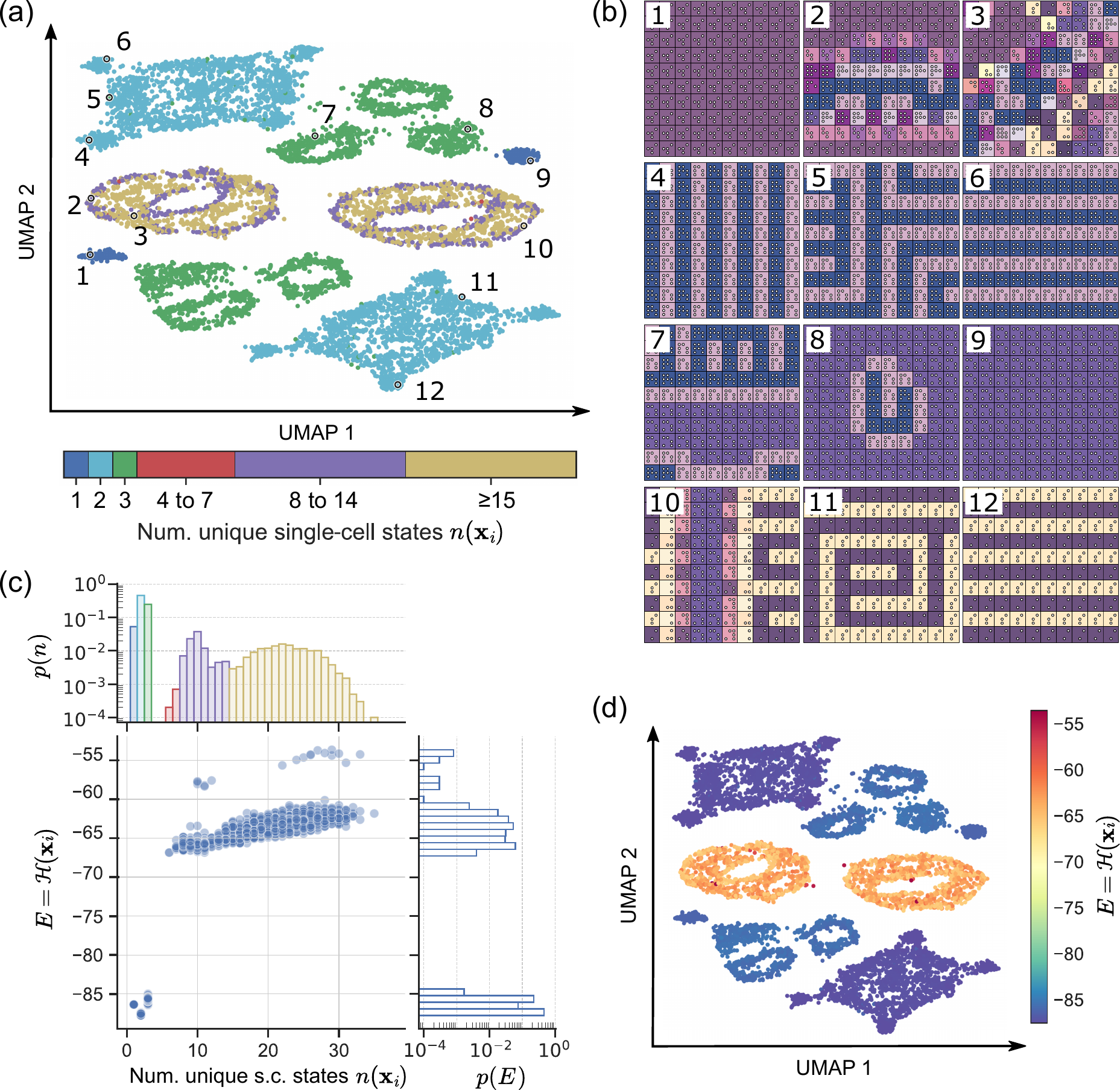}
  {Distribution of stable tissue configurations for strong signaling}
  {(a) The $\gamma=1$ sub-panel of Fig. \ref{multicell:fig3}(b) is shown with (b) twelve representative examples from the observed clusters (see Fig. \ref{multicell:fig2} for visualization details). 
  (c) For each point $\mathbf{x}_i$ from panel (a), the number of unique single-cell states present in the tissue, $n(\mathbf{x}_i)$, is plotted against the energy $E=H(\mathbf{x}_i)$ (Eq. (\ref{multicell:eq5})). 
  Histograms are shown for both distributions. 
  (d) The embedding from panel (a) is colored according to the multicell Hamiltonian $H(\mathbf{x}_i)$.}
  {multicell:fig4}{0.9\linewidth}

By comparing the annotated points within and between the clusters, we conclude that this unsupervised approach is producing intuitive clusters which group very similar tissue states (e.g. Fig. \ref{multicell:fig4}(b), examples $4, 5, 6$) while separating very different ones. In analogy to ``cell types", which are defined based on clusters of scRNA-seq data (and specified by gene expression signatures $\boldsymbol{\xi}^{\mu}$), we refer to these clusters as ``tissue types". In particular, we observe clusters of labyrinthine and stripe-like patterns wherein each cell is in one of two specific cell states (Fig. \ref{multicell:fig4}(b), examples $4-6, 11, 12$), clusters where the tissue gene expression is homogeneous (Fig. \ref{multicell:fig4}(b), examples $1$ and $9$), and clusters where two of the aforementioned tissue types are separated by an interface (Fig. \ref{multicell:fig4}(b), examples $2, 7, 8, 10$). 

We also note that within each cluster described above, there can be extensive fine-scale variation (for instance, in the spatial arrangement of cell states within the labyrinthine clusters). This suggests a hierarchical picture of the multicellular energy landscape Eq. (\ref{multicell:eq5}). At a high level, the landscape is partitioned into several basins of attraction corresponding to the different tissue types. In more detail, each of these basins may be locally very rugged (i.e. contain many local minima in close proximity), reflecting the large number of similar but distinct stable configurations we observe in certain clusters. 

Additionally, these clusters appear in symmetric pairs. Inspection of the elements of each pair reveals that they have opposite gene expression patterns: $\{ \mathbf{s}^a \} \rightarrow \{ -\mathbf{s}^a \}$, compare e.g. Fig. \ref{multicell:fig4}(b) points $6$ and $12$. 
This is a reflection of the spin-flip symmetry present in Eq. (\ref{multicell:eq5}), and indicates that the way the local minima are being sampled (i.e. the ensemble size and dynamical update rule) is capturing expected aspects of the energy landscape in a relatively unbiased manner. 
As an aside, the antisymmetric minima can be eliminated by applying an external field to the gene expression of each cell (e.g. by biasing certain ``housekeeping" genes to remain on), and this will be investigated in further work.

Separately, we also report the distribution of both $n(\mathbf{x}_i)$ and the multicellular energy $E=H(\mathbf{x}_i)$ over all sampled minima $\{ \mathbf{x}_i \}_{i=1}^k$ of $H(\mathbf{x})$. 
We display the data in Fig. \ref{multicell:fig4}(c), which contains a scatter plot as well as the two marginal distributions (i.e. $p(n)$ and $p(E)$). 
In Fig. \ref{multicell:fig4}(d) we provide a version of Fig. \ref{multicell:fig4}(a) colored instead by $E=H(\mathbf{x}_i)$. 

The data exhibits three main features. First, there is a clear correlation between $n(\mathbf{x}_i)$ and $H(\mathbf{x}_i)$ (note this is also apparent when comparing Fig. \ref{multicell:fig4} (a) and (d)). This means that the deepest minima tend to be compositionally simple, that is, characterized by a low number of unique single-cell states (low $n$ -- e.g. point $1$ in Fig. \ref{multicell:fig4}(b)). In contrast, the higher energy minima tend to be more complex, with large numbers of distinct single-cell states $n$, illustrated by point $3$ in Fig. \ref{multicell:fig4}(b). Second, these low energy, simple minima are also the most probable (over an ensemble of random initial conditions). This is reflected in the small $n$, low energy peaks in $p(n)$ and $p(E)$. In terms of the energy landscape, this suggests that the associated basins of attraction have large volume. And third, looking specifically at the distribution of the number of single-cell states within a tissue, $p(n)$, there is a smooth ``bulk" of compositionally complex minima with $5<n<35$ which appears bimodal. 

In the following subsection we investigate how these properties of the local minima of Eq. (\ref{multicell:eq5}) depend on the choice of random cell-cell signaling rules $\mathbf{W}$.

\subsection{Distribution of stable tissues under different random signaling rules displays universal characteristics}
\label{multicell:results_E}
As above we denote the frequency of local minima with $n$ unique single-cell states by $p(n)$, and the frequency of local minima with energy $E=H(\mathbf{x})$ by $p(E)$. We select eight different signaling rules $\{ \mathbf{W}_\alpha \}_{\alpha=1}^8$ with symmetric elements $W_{ij}=W_{ji} \sim U[-1,1]$. For each, we sample the local minima (self-organized tissue configurations) reached by an ensemble of random initial conditions. Each panel of Fig. \ref{multicell:fig5} corresponds to a particular $\mathbf{W}_\alpha$, and shows a scatter plot of $n(\mathbf{x}_i)$, $H(\mathbf{x}_i)$ for the sampled minima $\{ \mathbf{x}_i \}$ as well as the marginal distributions $p(n)$, $p(E)$. 

\newfigureWIDEtop{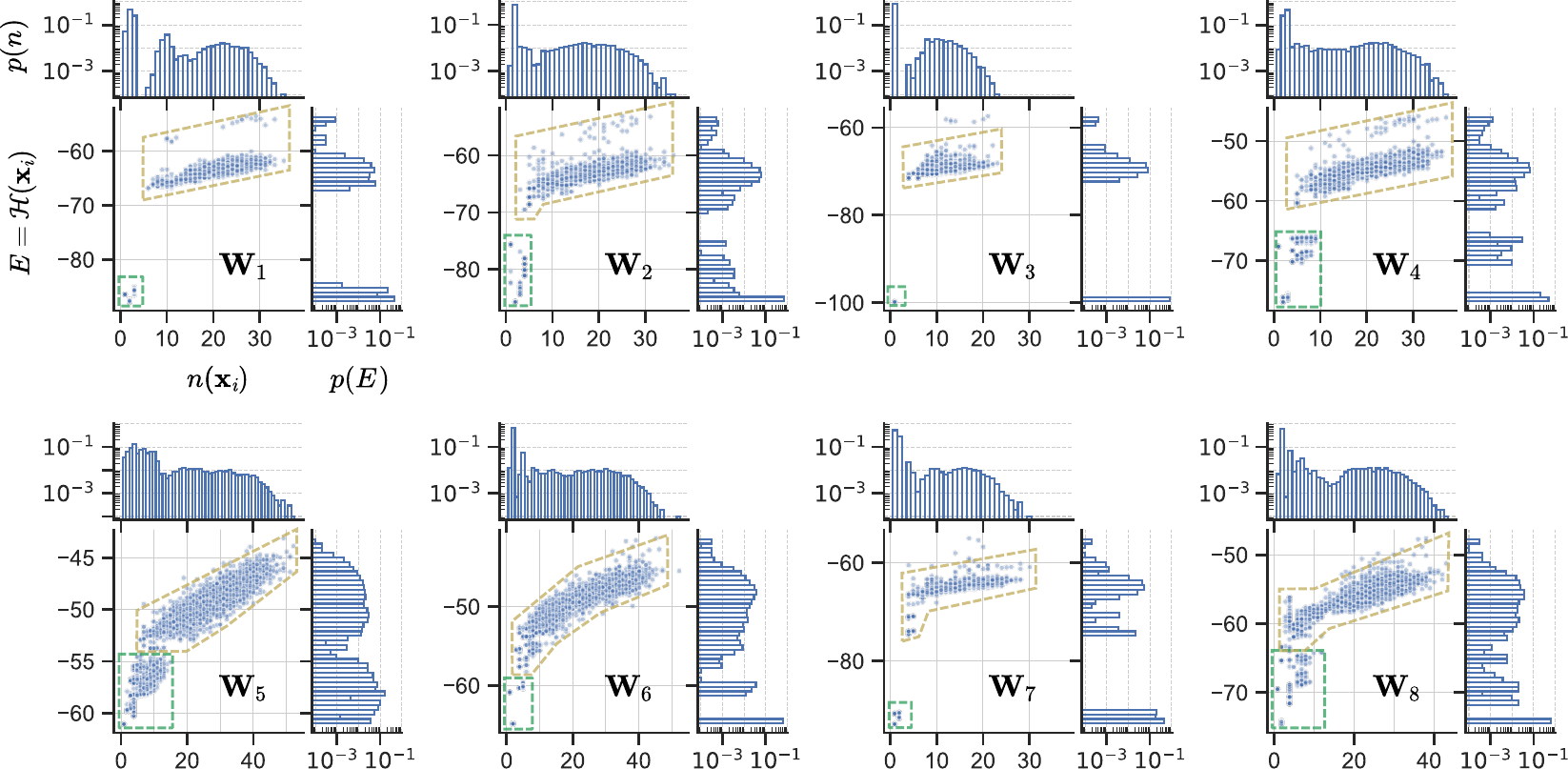}
  {Distribution of stable tissues under different signaling rules}
  {Eight signaling matrices $\{\mathbf{W}_{\alpha} \}$ are sampled according to $W_{ij} \sim U[-1,1]$. 
  For each $\mathbf{W}_{\alpha}$, the space of stable tissues (local minima of Eq. (\ref{multicell:eq5})) is sampled via $k=10^4$ random initial conditions at $\gamma =1$. 
  For each stable tissue $\mathbf{x}_i$ we compute its energy $E=H(\mathbf{x}_i)$ and the number of unique single-cell states $n(\mathbf{x}_i)$. 
  Each panel visualizes the joint distribution of these quantities as well as their marginal distributions $p(n)$, $p(E)$. 
  The dashed green (gold) box denotes the compositionally simple, low energy (complex, high energy) segments of the distribution. 
  The top-left panel $\mathbf{W}_1$ corresponds to the cell-cell signaling rules used in Figs. \ref{multicell:fig3} and \ref{multicell:fig4}. 
  All panels use the same $\mathbf{J}$, $\mathbf{A}$ as in Figs. \ref{multicell:fig3} and \ref{multicell:fig4}.}
  {multicell:fig5}{1.0\linewidth}

Despite some expected variation between the eight plots, the three features identified in the preceding subsection are largely preserved. All plots exhibit a positive correlation between the energy $E$ and the number of unique single-cell states $n$. This indicates that the deepest minima in the energy landscape Eq. (\ref{multicell:eq5}) are also the simplest (i.e. composed of few types of cells), which is denoted in Fig. \ref{multicell:fig5} by the dashed green box. Likewise, the more complex minima (large $n$) tend to have higher energies and form a ``bulk" (indicated by a dashed gold box in Fig. \ref{multicell:fig5}) which also has a positive slope. In all plots we see that these simpler, deep minima appear with much higher probability (note the logarithmic scale) than the more complex, shallow minima. 

This data suggests that, in the strong interaction regime, the energy landscape exhibits several universal features which are relatively invariant under different realizations of the cell-cell interaction matrix $\mathbf{W}$. Minor deviations arising from atypical $\mathbf{W}$ are observed, (e.g. the lack of a strong single peak at low $n$ for $\mathbf{W}_5$). Most notably, the lowest energy minima tend to also be the simplest minima, as quantified by $n(\mathbf{x})$. 
Furthermore, these deep, compositionally simple minima appear to have large basins of attraction when compared to the ``bulk" of more disordered, shallow minima. 

\section{Summary and Discussion}
\label{multicell:discussion}

Our work is motivated by a question that is central to the organization of multicellular life: how do tissues self-assemble into the numerous spatial architectures with diverse phenotypic compositions that make up multicellular organisms from the building blocks of genetically identical single cells? How can the cellular phenotypes on the one hand be sufficiently plastic and adaptable to generate this tissue diversity, while on the other remaining sufficiently stable to the dynamic signals found within different tissue microenvironments? While natural gene-gene interaction networks frequently appear disordered, with many weakly interacting genes and cross-wired signaling pathways, the resulting tissues are comparatively simple, often being composed of a handful of tissue-specific cell types that represent a small fraction of the possible phenotypic richness. The high-dimensionality of the interaction networks, in addition to their variability across different organisms, have posed longstanding challenges for studying questions concerning tissue self-organization and identifying general principles. 

Towards addressing these challenges, we have developed a generalized model of multicellular gene expression that couples single-cell gene regulation with cell-cell signaling in a manner that is scalable to large numbers of genes and cell types while being highly aligned to modern data modalities including scRNA-seq and spatial transcriptomics. 
Our model allows us to systematically study how tuning the degree of cell-to-cell coupling can control self-organization of multicellular collectives into different classes of stable tissue types with distinct spatial and compositional architecture. 
The basic unit of the model, the single cell, is parameterized by a set of gene expression vectors $\{ \boldsymbol{\xi}^{\mu} \}$ corresponding to single-cell types. These cell types are encoded as stable fixed points of a Hopfield network (HN), defining an effective gene-gene interaction matrix $\mathbf{J}$. 
In multicellular tissues described by a spatial adjacency matrix $\mathbf{A}$, genetically identical cells with the same internal rules $\mathbf{J}$ interact according to a cell-cell signaling matrix $\gamma \mathbf{W}$ which couples the gene expression of adjacent cells. 
For a given set of regulatory rules $(\mathbf{J},\mathbf{W})$, collective behavior emerges when the cell-cell signaling strength $\gamma$ is increased beyond a critical value $\gamma_c$, characterized by a cascade of collective transitions between different self-organizing tissue states which are generally unstable in the absence of cell-cell signaling. 
The punctuated nature of these transitions explains how flexible, phenotypically plastic cells can robustly assemble into a wealth of tissue configurations to carry out different functions when the strength of collective signaling is varied, as it does during development, wound healing, and the immune response. 

To emulate maximally disordered versions of the cross-wired signaling pathways often found in nature, we considered random cell-cell signaling networks $\mathbf{W}$. For strong signaling, where collective effects dominate, we found that disordered interactions tend to stabilize a surprisingly small number of tissue ``types" defined by large basins of attraction which can be locally rugged with many subtly distinct local minima. Interestingly, several features of the energy landscape appear invariant to different realizations of the random signaling rules. In particular, we find that minima depth and simplicity (as characterized by number of unique single-cell states and spatial structure) are positively correlated. Additionally, these deep, simple minima tend to have very large basins of attraction compared to the more complex, shallow minima. 
This contraction of the transcriptional landscape is consistent with the observation that multicellular tissues in different organisms generally contain only a small fraction of the possible phenotypic richness, often displaying just a handful of specialized cell types in each tissue, even when the microscopic interactions which drive the system -- defined here by the cell-cell signaling networks -- being highly disordered. 
These observations may have implications in other contexts, such as hierarchical self-organization in neuroscience \cite{Naim2018, OKane1992} as well as non-biological systems, for instance material self-assembly \cite{Murugan2015}.

Another important motivation for our work is the increasingly observed plasticity of cellular phenotypes across diverse tissue contexts. Certain classes of mammalian cells such as fibroblasts and macrophages exhibit tissue-specific variations in addition to shared subtypes found throughout the body \cite{Elmentaite2022}.
Tissue-specific cell type plasticity and abnormal changes in the cell type composition of tissue also play important roles in disease pathogenesis, including cancer initiation and maintenance \cite{Yuan2019, Giroux2017}. 
These observations of cross-tissue cell type heterogeneity raise numerous questions: When a cell type is ectopically placed in a ``sister" tissue rather than its native tissue, will it morph to its ``sister" cell type variant, and what compositional changes does one expect in the surrounding tissue? Such questions are difficult to address \textit{in vivo} or with existing modeling approaches but are highly suited to the model introduced here, which integrates the concept of cell type plasticity in diverse tissue contexts with tissue self-organization and multistability. 

Accordingly, our work has important implications for the definition of ``cell type". While the general concept is central to ongoing efforts to understand and ultimately treat disease by cataloging the gene expression of human tissues \cite{Karlsson2021, Regev2017}, the exact definition is still under debate \cite{Cell2017}. A common approach to define cell type is based on the identification of clusters of similar cells in single-cell gene expression data. 
A more nuanced view is suggested by our framework, which reveals how the fixed points of the multicellular system (stable cell states within tissue) evolve with the strength of the intercellular signaling $\gamma$. 
At low $\gamma$ the prescribed single-cell types remain stable, but as $\gamma$ increases they can morph as multicellular structures progressively arise. 
\textit{In vivo}, cells are continuously interacting and these interactions control the stability of different cell states in tissue. This effect -- which should be taken into account by emerging definitions of cell type -- is formalized within our model. By disentangling the intra- and intercellular levels of gene regulation, our work provides a quantitative picture of cell type plasticity in which gene regulatory fixed points shift and bifurcate upon exposure to different tissue microenvironments. 

Recent gene-free approaches which model the effects of cell-cell signaling on a low-dimensional phenotype space \cite{Camacho-Aguilar2021, Corson2017, Corson2012, Corson2017a, Rand2021, Saez2021} provide an elegant method for describing cell fate transitions. 
Such approaches rely on bifurcation analysis of proximal cell types, and thus far have been limited to a handful of related cell types represented as phenotypic fixed points. Our framework is able to accommodate many disparate cell and tissue types while using a less abstract (but much more detailed) state space that represents the expression state of individual genes in each cell. Towards bridging the two modeling techniques, we note that the presented model is naturally equipped with a low-dimensional phenotype space: the projection of the single-cell states onto the embedded cell types \cite{Amit1989, Kanter1987, Kramer2021, Lang2014, Pusuluri2018}. 
This may be an avenue to develop a theoretical description of cell fate transitions in our framework while improving the interpretability of the low-dimensional phenotype space. 

The current theoretical work provides the basis for future applications to specific experimental systems. 
Inspired by the past successes of other modeling approaches \cite{Camacho-Aguilar2021, Corson2012, Corson2017}, this will involve experimentally informed (non-random) $\mathbf{W}$, tissue-specific choices of adjacency matrix $\mathbf{A}$, and gene expression noise. 
More broadly, our framework is designed with emerging sequencing techniques in mind, specifically spatial transcriptomics \cite{Longo2021}. 
By using many snapshots of gene expression in tissue, $\mathbf{J}$, $\mathbf{W}$, $\mathbf{A}$, and $\gamma$ can be conditionally inferred (or at least constrained). 
Towards overcoming potential inference challenges related to undersampling, we note that the present model and future generalizations may be represented as graph neural networks (GNNs) \cite{Scarselli2009, Zhou2020}, and the rapidly progressing GNN literature may reveal efficient inference techniques. 
Future work in this direction will serve as an important test of the framework’s experimental predictions. 

There are several potential next steps to deepen the theoretical understanding of the presented framework and extend its applicability to broader biological contexts. First, a theoretical explanation should be developed to understand the rich sequence of tissue transitions at intermediate values of $\gamma$ (which will depend on the detailed structure of $\mathbf{J}$ and $\mathbf{W}$), as well as the invariant statistical properties for the minima distributions in the high-$\gamma$ limit. 
This will serve to generalize our results which are based on finite systems with $N=9$ genes and $p=3$ encoded cell types. 
Relatedly, it would be useful for engineering purposes to be able to predict the types of tissues than can self-organize (e.g. labyrinthine or homogeneous) given arbitrary $\mathbf{J}$, $\mathbf{W}$, $\mathbf{A}$, and $\gamma$. 
Second, for the random cell-cell signaling we have considered, the stable tissue phenotypes at high $\gamma$ do not necessarily contain the encoded cell types; the case of non-random cell-cell interactions matrices $\mathbf{W}$ which preserve the encoded single-cell types $\{ \boldsymbol{\xi}^{\mu} \}$ will be investigated to better understand cellular plasticity within tissue. 
Finally, cellular division and apoptosis events occur when particular gene expression patterns are reached; incorporating these events in the model will provide a natural framework to study the sequences of self-organized tissue transitions which characterize development \cite{Stanoev2021}.
The intra- and intercellular regulatory rules $\mathbf{J}$, $\mathbf{W}$ will determine the fate of these unfolding trajectories.

\begin{acknowledgments}
We acknowledge the support of the Natural Sciences and Engineering Research Council of Canada (NSERC) through Discovery Grant RGPIN
402591 to A.Z.; CGS-D Graduate Fellowship to M.S. 
We thank Duncan Kirby and Jeremy Rothschild for useful discussions, and Tatyana Gavrilchenko and Pearson Miller for helpful comments on the manuscript. 
\end{acknowledgments}

\input{main.bbl}                 

\end{document}

%% file: main.bbl
%